\documentstyle[12pt,epsf]{article}
\def\ga{\mathrel{\raise.3ex\hbox{$>$\kern-.75em\lower1ex\hbox{$\sim$}}}}
\def\la{\mathrel{\raise.3ex\hbox{$<$\kern-.75em\lower1ex\hbox{$\sim$}}}}
\def\gev{{\rm \, Ge\kern-0.125em V}}
\def\tev{{\rm \, Te\kern-0.125em V}}
\def\beq{\begin{equation}}
\def\eeq{\end{equation}}

\def\ss{\scriptscriptstyle}

\def\mpar{m_{\ss\|}^2}
\def\mpl{M_{\rm Pl}}
\def\p{\varphi}

\def\Tc{T_{\rm crit}}
\def\mpar{m_{\ss ||}}
\def\mave{\hat m_{\tilde q}}

\begin{document}
\begin{titlepage}
\pagestyle{empty}
\baselineskip=21pt
\rightline{hep-ph/9611325}
\rightline{UMN--TH--1518/96}
\rightline{TPI--MINN--96/20}
\rightline{LANCS-TH/96/15}
\rightline{UCSBTH--96--28}

\rightline{November 1996}
\vskip1.25in
\begin{center}
  {\large{\bf Constraints from Inflation and Reheating on Superpartner Masses}}
\end{center}
\begin{center}
\vskip 0.5in
{Toby Falk,$^1$ Keith A.~Olive,$^1$  Leszek Roszkowski,$^2$
Anupam Singh,$^3$ and Mark Srednicki$^3$
}\\
\vskip 0.25in
{\it
$^1${School of Physics and Astronomy,
University of Minnesota, Minneapolis, MN 55455, USA}\\
$^2${School of Physics and Chemistry, Lancaster University,
Lancaster LA1 4YB, Great Britain}\\
$^3${Department of Physics,
University of California, Santa Barbara, CA 93106, USA}\\}
\vskip 0.5in
{\bf Abstract}
\end{center}
\baselineskip=18pt \noindent
%%%%%%%%%%%%%%%%%%%%%%%%%%%%%%%%%%%%%%%%%%%%%%%%%%%%%%%%%%%%%%%%%%%%%
Flat directions in the Minimal Supersymmetric Standard Model (MSSM)
can become unstable due to radiative corrections, and the global
minimum of the (zero temperature) potential can lie at large values of
the squark and/or slepton fields.  Here we show that, in inflationary
models of early cosmology, the universe is very likely to be in the
domain of attraction of this global minimum at the end of inflation.
While the minimum at the origin of field space is global at
sufficiently high temperatures, depending on details of the model, the
universe may be trapped in the non-zero minimum until it becomes the
global minimum at low temperatures.  Parameter values leading to this
scenario are therefore ruled out.
%%%%%%%%%%%%%%%%%%%%%%%%%%%%%%%%%%%%%%%%%%%%%%%%%%%%%%%%%%%%%%%%%%%%%
\end{titlepage}
%\newpage
\baselineskip=18pt
%%%%%%%%%%%%%%%%%%%%%%%%%%%%%%%%%%%%%%%%%%%%%%%%%%%%%%%%%%%%%%%%%%%%%
In an earlier paper \cite{us}, we pointed out that certain mass
patterns for the superpartners result in a radiatively-corrected
scalar potential whose global minimum is at large values of the squark
and/or slepton fields.  Because these minima are in general electric
charge and color breaking, we concluded that these mass patterns were
therefore disallowed, or at least required new physics beyond the
standard model.  However, Riotto and Roulet \cite{rr} argued that the
desired minimum (with zero squark and slepton VEVs and the usual
nonzero Higgs VEVs), while not global at zero temperature, is
generically the global minimum at high temperature.  Furthermore, if,
for some reason, the universe begins in the domain of attraction of
this minimum, then it will get stuck there, since the probability to
tunnel to the global minimum (after the temperature drops
sufficiently) is much too small.  A similar argument was made in
\cite{kls}.

The assumption made (either explicitly \cite{rr} or implicitly \cite{kls})
is that the universe will begin in the global minimum of the temperature
dependent scalar potential.  This, however, is extremely unlikely
to be the case along flat directions in field space
after a period of inflation \cite{inf}.
During inflation, fluctuations in scalar fields which parameterize flat
directions, i.e. those with supersymmetry breaking squared masses $m^2 \ll H^2$
increase with time according to \cite{lin2}
\begin{equation}
\langle\p^2\rangle = {1\over 4\pi^2}H^3\tau \; ,
\label{des}
\end{equation}
where $H$ is the Hubble constant during the inflationary epoch.
Subsequently, the long wavelength modes of these fluctuations behave
as a classical background field with amplitude $\p_0 \simeq
{\langle\p^2\rangle}^{1/2}$. More precisely, the value of $\p_0$ we
compute in this way is the RMS width of a probability distribution
(which is approximately Gaussian) for the value of $\p$ in a horizon
volume.  Solving the horizon and flatness problems requires that the
duration $\tau$ of the inflationary era satisfies $H\tau \ga 60$.  In
very general models of inflation which involve a single mass scale
$m_I$, we have $H\tau \simeq (M_P/m_I)^2$, where $M_P$ is the Planck
mass.  The value of $m_I$ determines the size of the observed
fluctuations in the microwave background, and their measured value
fixes $(m_I/M_P)^2 \simeq \hbox{few} \times 10^{-8}$ \cite{cdo}, which
then implies $H\tau \ga 10^8$ \cite{eeno}.  The case of interest for
us will be that of a squark or slepton field $\p$ (which is not the
inflaton field) with a slowly varying potential $V(\p)$ corresponding
to an almost flat direction.  We define the field-dependent squared
mass $m^2(\p)=(1/2)\partial^2 V(\p)/\partial\p^2$.  The field stops
evolving according to eq.(\ref{des}) when $m^2(\p) t \simeq 3 H / 2$.
Then we find \cite{lin2}
\begin{equation}
(\p_0^{\rm max})^2 = \min\left\{{1\over 4\pi^2}H^3\tau, \ {3\over 8\pi^2}
{H^4\over m^2(\p)}\right\} \; .
\label{pmax}
\end{equation}
For $H^2 \simeq m_I^4/M_P^2 \simeq 10^{-15} M_P^2$, and for (nearly)
flat directions characterized by a mass scale $m(\p) \simeq 10^{-16}
M_P$, the first term in eq. (\ref{pmax}) is smaller than the second,
and the growth of fluctuations is therefore limited by the duration
$\tau$ of the inflationary era.  This generally results in a value of
$\p_0$ which is much larger than $H$, $\p_0 \simeq 10^{-4} M_P$.

Thus, if $\p_0$ is as large or larger than the location of the global
minimum of $V(\p)$, we are overwhelmingly likely to find ourselves,
after inflation, in the domain of attraction of this minimum.  This
fact (that scalar fields along flat directions find themselves far
from the origin in field space after inflation) has long been
recognized and is the basis of the Affleck-Dine mechanism of
baryogenesis \cite{ad,lin3,eeno}, which requires large initial values
of squark and slepton fields.  This scenario is particularly
attractive because one can in general find flat directions in field
space in supersymmetric grand unified theories.  That is, these
theories naturally contain directions in which the scalar potential is
absolutely flat $V(\p) = 0$ up to supersymmetry breaking effects which
induce masses of order the supersymmetry breaking scale. $\p$ is some
combination of squark and slepton and Higgs fields in which the F- and
D- terms in the scalar potential vanish.  If the expectation value of
$\p$ is non-zero (and large) initially (e.g. due to inflation as
described above), the subsequent evolution of certain flat direction
in a supersymmetric GUT can be shown to give rise to a baryon
asymmetry.

It has recently been pointed out however, that the simple picture of
driving scalar fields to large vacuum values along flat directions
during inflation is dramatically altered in the context of
supergravity \cite{drt}.  During inflation, the Universe is dominated
by the vacuum energy density, $V \sim H^2 M_P^2$. The presence of a
non-vanishing and positive vacuum energy density indicates that
supergravity is broken and soft masses of order of $H$ are
generated \cite{all}.  The implications of such terms for the charge
and color breaking minima and constraints on the MSSM parameters have
also been recently addressed \cite{strum}.  In minimal supergravity it
is quite easy to see that such mass terms are generated. In general,
the scalar potential in a supergravity model is described by a
K\"ahler potential $G$ \cite{sugr} and in minimal supergravity we
define the K\"ahler potential by
\begin{equation}
G = \phi_i^* \phi^i + \ln |W(\phi)|^2
\label{min}
\end{equation}
where $\phi^i$ represents all scalar fields in the theory and
$W(\phi)$ is the superpotential.  This results in a scalar potential
of the form
\begin{equation}
  V = e^G \left[ |W_{i} + \phi^*_i W|^2/|W|^2 - 3 \right]
\label{gen}
\end{equation}
where $W_i = \partial G/\partial \phi^i$.  A positive vacuum energy
density, $V > 0$, needed for inflation, breaks supergravity and both
the exponential and the term enclosed in brackets in (\ref{gen}) must
be non-vanishing.  In fact, the exponential $e^G$ is the order
parameter for supergravity breaking.  Included among the $\phi^i$'s
is the flat direction $\p$, and by inspection of (\ref{gen}) one
finds a mass term $e^G \p \p^*$. A large mass term of this type
precludes the possibility of Affleck-Dine baryogenesis.  It was noted
in \cite{drt} that A-D baryogenesis could be revived in non-minimal
models in which such a mass terms arises with an opposite sign having
the effect of driving $\p$ to extremely large VEVs.  In fact, it was
shown in \cite{gmo}, that in supergravity models which possess a
Heisenberg symmetry \cite{BG}, including no-scale models of
supergravity \cite{ns,nsguts}, supersymmetry breaking makes no
contribution to scalar masses, leaving supersymmetric flat directions
flat at tree-level.  One-loop corrections in general lift the flat
directions, but naturally give small negative squared masses $\sim -
g^2 H^2/(4\pi)^2$ for all flat directions that do not involve the
stop. In the context of these theories, we therefore expect initial
field values for $\p$ to be quite large.

In the types of models discussed above, reheating is generally quite
inefficient.  In models in which the inflation is coupled only gravitationally
to the observable (gauged) sector, the reheating temperature
is in fact very low \cite{eeno},
$T_R \sim \alpha^2 m_I^3/M_P^2 \sim 10^5\gev$.
However, there has been a great deal of recent activity and progress in
understanding the quantitative details of the process of particle
production after cosmological phase transitions and the implications of this for
reheating after inflation \cite{linde1,linde2,partlist,dissip,prestable},
and we would like to briefly discuss the relevant aspects of these
developments so as to address their implications for our work here.
One of the key issues is whether or not
there is symmetry restoration due to the effects of quantum
fluctuations \cite{linde2}.  It is now clear that the occurrence of this 
phenomenon is model dependent \cite{linde2,dissip,prestable}. 
It has been pointed out \cite{prestable} that symmetry restoration effects
may help localize the fields we are interested in close to the origin.
However, these effects can occur in the models of interest
only if specific couplings between
the inflaton, an additional scalar field, and the flat-direction field $\varphi$
are present and assumed to be large \cite{prestable}.
In a general model, particularly one in which the inflaton
has only gravitational interactions with other fields, symmetry restoration
is not expected to occur.

In this paper we wish to determine the conditions under which gauge
symmetries remain broken after inflation due to the presence of
additional minima along flat directions.  In the problem we are
interested in, although for sufficiently high reheat temperatures the
minimum at $\p = 0$ is a global minimum subsequent to the reheating of
the universe after inflation, there is another minimum of the
potential at $\p \ne 0$ which becomes the global minimum at lower
temperatures and hence later in the evolution of the universe.
During the epoch when the $\p \ne 0$ minimum is the
metastable minimum of the potential, it is a relevant and appropriate
question to ask whether the tunneling rate is such that we will end up
at $\p = 0$ due to tunneling. We will examine this rate and hence the
possibility of such a tunneling in this paper.

At tree level,
the zero temperature potential takes the form
\begin{equation}
V(\p,Q)=\frac12 m^2(Q)\p^2 + \frac1n M^{4-n}\p^n \;,
\label{pot}
\end{equation}
where the mass-squared $m^2(Q)$ depends on the renormalization group
scale $Q$, and $M$ is a new mass scale (such as the Planck mass) which
characterizes the strength of non-renormalizable interactions due to
physics beyond the minimal supersymmetric standard model (MSSM).  Loop
corrections, if included to all orders, would remove the $Q$
dependence of $V$.  We will work with the RG-improved tree-level potential,
but attempt to optimize the choice of $Q$ so as to minimize the
contributions of the loop corrections.
We treat the case of flat directions, where there is no
renormalizable interaction term.  The lowest value of $n$
which is allowed depends on the flat direction of interest \cite{gher};
for the case
$\tilde u_R^{\ss r} = \tilde s_R^{\ss g} = \tilde b_R^{\ss b} \equiv v(Q)$
which we treated earlier, we have $n_{\rm min}=10$.  For the case
$\tilde u_R^{\ss r} = \tilde c_R^{\ss g} = \tilde s_R^{\ss b}
= \tilde e_R \equiv v(Q)$,
we have $n_{\rm min}=6$.
In our numerical investigation of the tunneling rates,
we concentrate on the case $n_{\rm min}=6$, as
the rates will be even lower in the case $n_{\rm min}=10$.
For both flat directions, the main contribution to the running of
$m^2(Q)$ comes from SU(3) interactions, which result in \cite{us}
\begin{equation}
m^2(Q)=\mpar^2
- {2\over\pi^2} g_3^2(M_3)M_3^2 \ln(Q/M_3)
\left\{{1+3g_3^2(M_3)\ln(Q/M_3)/16\pi^2 \over
\left[1+3g_3^2(M_3)\ln(Q/M_3)/8\pi^2\right]^2} \right\}.
\label{rgesoln}
\end{equation}
where $M_3$ is the gluino mass at the scale $Q=M_3$, and $\mpar^2$ is
the sum of the squark squared masses in the flat direction at a scale
$M_3$.  We will ignore the effects of the electroweak gauge couplings
and the Yukawa couplings, which are sub-dominant.  At large enough $Q$,
$m^2(Q)$ becomes negative, and the minimum of the effective potential
occurs at $\p=v(Q)$ where
\begin{equation}
v(Q) = M^{(n-4)/(n-2)}[-m^2(Q)]^{1/(n-2)}.
\label{vq}
\end{equation}
The best value of $Q$ to choose (that is, the value which is likely to
minimize the higher-loop contributions near the minimum) is the
solution (if it exists) of $Q=v(Q)$; if no solution exists, the
apparent minimum is likely to be an invalid artifact of the one-loop
approximation.  Alternatively, we can just choose $Q=\p$ and find the
minimum of $V(\p,\p)$.  A slightly better choice is \cite{rr}
$Q=(\p^2+M^2_3)^{1/2}$, which prevents problems at small values of
$\p$, and this is the scheme we adopt.

The temperature-dependent corrections to $V(Q,\p)$ take the form \cite{dj}
\begin{equation}
V_T(\p)={T^4\over2\pi^2}\sum_i\epsilon_i\int_0^\infty
dq\,q^2\ln\left[1-\epsilon_i\exp\left(-[q^2+m_i^2(\p)/T^2]^{1/2}\right)\right],
\label{vtf}
\end{equation}
where $T$ is the temperature, the sum is over all particle species
whose tree-level masses $m_i$ depend on $\p$, and $\epsilon_i = +1$
for a boson and $-1$ for a fermion.  For the flat directions we
consider, there are a total of 32 particles which couple strongly
  to the flat direction that contribute to the $\p$ dependence of
$V_T(\p)$: 8 gluons with mass $g_3\p$, 8 squarks with mass $(m_0^2 +
g^2_3\p^2)^{1/2}$, 8 gluinos with mass $\frac12[(M_3^2 +
2g^2_3\p^2)^{1/2}+M_3]$, and 8 quarks with mass $\frac12[(M_3^2 +
2g^2_3\p^2)^{1/2}-M_3]$.  (Actually, the fermion mass eigenstates are
quark/gluino mixtures.)  The effects of $V_T(\p)$ are significant for
$g_3\, \p \la T$, where it can be approximated as
\begin{equation}
V_T(\p) \simeq -{\textstyle{2\over3}}\pi^2 T^4 +2 g_3^2 T^2 \p^2.
\label{vt}
\end{equation}
There are also particles which couple to the flat directions weakly
and via Yukawa interactions, and these particles acquire much smaller
masses when $\p$ is large than those which couple strongly to $\p$.
Thus when $\p\ga T/g_3$, the number density of the strongly coupled
fields becomes exponentially suppressed, but the weakly coupled fields
can still contribute to $V_T$, and it is therefore these weakly
coupled degrees of freedom which are numerically important in
determining the temperature at which these large scale minima
disappear\footnote{We thank Alessandro Strumia for pointing this out to us.}. 
 These light degrees of freedom include
bino/quark,  hypercharge gauge boson/squark, Higgsino/quark and
Higgs/squark admixtures, and for sufficiently high temperatures, they
contribute an amount
\begin{equation}
  \label{vlight}
  \Delta V_T(\p) = ( {\textstyle{1\over4}}c_n^2 g_1^2 + 
{g_2^2\over 4
    M_W^2 \sin^2 \beta} m_+^2 + {g_2^2\over 4 M_W^2\cos^2 \beta} m_-^2)
 T^2 \p^2
\end{equation}
to the effective potential, where $c_6(c_{10}) = \sqrt{20\over
  9}(0)$ and $\{m_+,m_-\} = \{m_c,m_s\}(\{m_u,m_b\})$ for $n_{\rm
    min} = 6(10)$.  In the $n = 10$ case, $c_{10} = 0$ because
there remains an 
unbroken U(1) gauge symmetry which is a linear combination of 
hypercharge and a color generator.
For flat directions corresponding to $n_{\rm min}=6$, the
second minimum of $V_{T}$ (when it exists) occurs at $v\sim
10^{11}\gev$, for $\sim\tev$ gluino masses.  For $n_{\rm min}=10$,
$v\sim 10^{15}\gev$.

To calculate the tunneling rate from the $\p\neq 0$ minimum to the
origin , we need to solve the equation
\begin{equation}
\left(\partial_r^2 +{2\over r}\partial_r\right)\p = {\partial\over\partial\p}
\left[ V(Q,\p) + V_T(\p)\right]_{Q=(\p+M_3^2)^{1/2}}
\label{bounce}
\end{equation}
subject to the boundary conditions $\partial_r\p|_{r=0}=0$ and
$\p(\infty)=v$, where $v$ is the location of the minimum of
$V(Q,\p)+V_T(\p)$.
The tunneling rate is then proportional to $\exp(-S_3/T)$, where
the three-dimensional action is given by
\begin{equation}
S_3 = 4\pi\int_0^\infty dr\,r^2\left[{\textstyle{1\over2}}(\partial_r\p)^2
+ V(Q,\p) + V_T(\p) - V_0\right],
\label{s3}
\end{equation}
where $V_0$ is a constant which cancels off the potential energy at $\p=v$.
$S_3$ can be computed numerically, but to gain intuition we note that
at small $\p$ the potential is dominated by $V_T(\p)$; the desired solution
is then
\begin{equation}
\p(r) = A{\sinh(2 g_3 T r)\over r}
\end{equation}
where $A$ is a constant.  Near the minimum, the potential can be
approximated as
\begin{equation}
V(\p) \simeq {\textstyle{1\over2}}m^2(\p-v)^2
\label{vmin}
\end{equation}
where $m^2 \sim |m^2(v)|$ is positive.  Near $\p=v$ the desired solution is
\begin{equation}
\p = v - B {\exp(-mr)\over r}
\end{equation}
where $B$ is another constant.  If we now make the drastic approximation
that the full potential is given by eq.(\ref{vt}) for $\p<\p_c$ and
by eq.(\ref{vmin}) for $\p>\p_c$, with $\p_c$ given by the value of $\p$
where the two approximate potentials equal each other, then it is possible
to solve for $A$, $B$, and $S_3$ analytically.  For $T \la v$ and
$m \ll T^2/v$, we find $S_3/T \sim v^3/T^3$.
Thus we must have $T$ near $v$ in order to have a fast enough tunneling
rate.

On the other hand, for $T$ near $v$, the field-dependent part of $V_T$
dominates the zero-temperature potential at $\p=v$, and for $T$ larger
than some critical temperature $\Tc(\mpar, M_3)$, the second minimum
(corresponding to non-zero $v$) along the flat direction is removed.
In fig.~1, we plot contours of $\Tc (\mpar, M_3)$, in units of
$10^7\gev$, for the $n_{\rm min}=6$ flat direction $\tilde
u_R^{\ss r} = \tilde c_R^{\ss g} = \tilde s_R^{\ss b} = \tilde e_R
\equiv v(Q)$.  We have taken $\tan \beta = 2$. 
The $x$-axis specifies the gluino mass $M_3(M_3)$,
while the $y$-axis labels $\mave(\mave)$, where
$\mave^2=(\mpar^2-m_{\tilde e_{\ss R}}^2)/3$ is the average squark
mass$^2$ in the flat direction for $m_{\tilde e_R}^2\ll m_{\tilde q}$.
Above the solid curve, which lies along the line $\mave\simeq 0.76
M_3$ for large $M_3$, the zero temperature potential $V(\p,Q)$ (and
hence the full potential $V(\p,Q) +V_T$) has no second minimum ($\Tc$
drops off very rapidly to zero above the top contour).  We see that if
the reheating temperature $T_R$ is $\la 10^7\gev$, the
finite temperature effective potential after reheating will contain
large scale minima for the same parameter values as for the
zero-temperature effective potential.  For larger $n_{\rm min}$, $v$
is larger, along with the correspondingly larger critical
temperatures.  In fig.~2, we plot contours of $\Tc (\mpar, M_3)$, in
units of $10^9\gev$, for the $n_{\rm min}=10$ flat direction $\tilde
u_R^{\ss r} = \tilde s_R^{\ss g} = \tilde b_R^{\ss b} \equiv v(Q)$,
and we see that for this direction we require larger reheat
temperatures, on the order of a few $\times 10^9\gev$, in order to
remove the $\p\ne 0$ minima.  In this case the average squark mass
$\mave$ is defined by $\mave^2=\mpar^2/3$.

After inflation, the field $\p$ parameterizing the flat direction
rolls from $\p_0$, given by (\ref{pmax}), toward the the nearest
minimum of the potential.  Because the time-scale for the field to
reach this minimum is much shorter than the time for decay of the
inflaton and the subsequent thermalization of its decay products
(assuming, as above, that the inflaton is gravitationally coupled to
ordinary matter), the evolution of $\p$ is determined by the
zero-temperature effective potential.  The initial conditions at
reheating are therefore determined by the value of $\p_0$ relative to
the maximum of $V$ separating the $\p = 0$ and $\p \ne 0$ minima.  If
it is to the left of the barrier, $\p$ will roll to the origin and
remain there and we recover the scenario described by \cite{rr}.  If
$\p_0$ is to the right of the barrier, it will roll to the $\p \ne 0$
minimum (unless it is so far to the right that it picks up a
sufficient velocity so as to carry it over the barrier and settle once
again at the origin).  When non-renormalizable interactions for flat
directions are included, the maximum value for $\p_0$ is limited by
the effective mass at large $\p$.  For our $n=6$ case, the minimum is
at roughly $\p \sim 10^{11}\gev$, and the position of the barrier
depends on $\mpar$ and $M_3$, but is typically much smaller than $v$,
while ${\p_0^{\rm max}} \la 10^{14}\gev$.  Thus we expect to be to the
right of the barrier, and we have determined that so long as $\p_0 <
1.5 \times 10^{14}\gev$, the field will settle in the broken vacuum.
For $n=10$, the minimum is shifted up to about $\p \sim 10^{15} \gev$
and is comparable to the maximum value of $\p_0$, while again the
maximum of the barrier is a much lower scales.  Thus if the {\em
  zero-temperature} potential for $\p$ has a large-scale minimum, this
is where we expect the field to sit at the beginning of reheating.  If
$T_R<\Tc(\mpar, M_3)$, the position of the minimum of $V_{\rm eff}$ is
unaffected by the temperature corrections, and after reheating $\p$
will be trapped in the symmetry-breaking minimum.

It is possible that thermal effects can excite $\p$ over the potential
barrier and allow nucleation of bubbles of the symmetric vacuum.
In order to determine whether $\p$ remains trapped in the $\p\ne0$
minimum, we numerically compute the finite temperature transition rate
from the $\p\ne0$ minimum to the true minimum (of the
finite-temperature potential) at $\p=0$.  For
fixed temperature $T$, we use the full finite-temperature corrections
(\ref{vtf}) to the effective potential to compute the bounce solution
to (\ref{bounce}) and the resulting three-dimensional action (\ref{s3}).
The tunneling probability per volume per unit time is \cite{linde}
\begin{equation}
  \label{gamma}
\Gamma\sim T^4 e^{-S_3/T}.
\end{equation}
The fraction of space remaining in the broken phase is $f=e^{-P}$ \cite{gw},
where
\begin{equation}
  \label{prob}
  P\,\sim\,{\mpl^4\over T_0^3} \int^{T_R}_{T_d} T^{-2}
       \left(1-{T_0\over T}\right)^3 {e^{-S_3/T}} dT \,<\,
       {\mpl^4\over T_d T_0^3}\; e^{-S_3/T_R},
\end{equation}
and where $T_0$ is the current temperature of the universe and $T_d$
is the temperature at which the two minima become degenerate.  For
$n_{\rm min}=6$ flat directions, $T_d$ is of order $10^6-10^7\gev$,
while for $n_{\rm min}=10$ flat directions $T_d$ is of order
$10^8-10^9\gev$; for reheat temperatures less than this, there is of
course no tunneling. The expression on the right-hand side of
(\ref{prob}) can be a gross overestimate, depending on how fast $S_3$
drops off as the temperature falls.  As $P$ is exponentially sensitive
to $S_3$, we can take $P=1$ as a critical value, such that $P<1$
implies no nucleation of the symmetric phase.  We then find that $\p$
remains trapped in the large-scale minimum as long as
\begin{equation}
  \label{s3bound}
{S_3\over T_R} >217 + \ln\left({10^9\gev\over T_R}\right)
       + \ln\left({T_R\over T_d}\right)
\end{equation}
For fixed values of $(\mpar, M_3)$, we have looked for values of
$T_R$ satisfying $T_R<\Tc(\mpar, M_3)$, but violating (\ref{s3bound}).
We find that $S_3$ rises very quickly as $T_R$ drops below $\Tc$, and
we were unable to resolve a single case where (\ref{s3bound}) does not
hold by better than an order of magnitude.  That is, $T_R$ must be
extremely close to $\Tc$ in order for (\ref{s3bound}) to be violated
(in which case the integral in (\ref{prob}) would need to be evaluated
numerically anyway, yielding a much lower value for $P$ than given by
the right-hand-side of (\ref{prob})).  (In \cite{rr}, tunneling from
$\p=0$ to $\p=v$ was found to occur over non-negligible regions of the
parameter space at small $\mpar$, where the width of the barrier was
small ($<1\tev$); in this case, the barrier widths are much larger).
We can therefore say that {\em if the reheat temperature $T_R$ is less
  than the critical temperature $\Tc$ (shown in Figure~(1)), we expect
  the field $\p$ parameterizing the flat direction to become trapped
  in the large-scale symmetry-breaking minimum, and to remain there
  until the large-scale minimum becomes the global minimum at low
  temperatures}.  This effectively rules out regions of $(\mave, M_3)$
parameter space where such minima exist, or at the very least implies
new physics beyond the standard model.

{}From fig.~2, we see that typical values of $T_{\rm crit}$ for the
$n_{\rm min}=10$ flat direction are about a few times $10^9\gev$. We
can ask whether reheat temperatures of this magnitude are reasonable.
There are of course strong limits on the reheat temperature in
supersymmetric theories due to gravitino production. If the gravitino
mass is related to the supersymmetry breaking scale (though this need
not be the case in no-scale supergravity \cite{grns}) and $m_{3/2}
\sim 1\tev$, then the production of gravitinos during the reheat
process and their subsequent decay will lead to the photodestruction
of deuterium and $^4$He.  This allows one \cite{gr} to set a limit on
the reheat temperature which is comparable to $T_{\rm crit}$, i.e.
$T_R \la$ a few $\times 10^9\gev$, though this limit can be evaded for
a significantly larger gravitino mass.  We note that the gravitino
decay bound was re-examined in the light of parametric resonance by
Allahverdi and Campbell \cite{ac}. They concluded that these bounds
cannot be evaded even in the presence of parametric resonance.

In summary, we find that after an inflationary epoch, fields
parameterizing flat directions in the SUSY scalar potential will
settle into large-scale ($>10^{10}\gev$) minima, if they exist, and
will remain there through reheating until the minima become global
minima at low temperatures.  This effectively confirms the qualitative
conclusions of \cite{us}, albeit with some new model-dependence.  In
particular, we conclude that the SUSY particle mass relationships
corresponding to the region under the solid line in fig.~1 are
inconsistent with inflationary models, such as those in no-scale
supergravity, in which 1) the inflaton is gravitationally coupled to
the observable sector,
or in any model in which the reheat temperature is less than that given by
the contours in fig.~1 or fig.~2, depending on which flat direction is
populated during inflation, and 2) there are no positive $O(H^2)$
contributions to the scalar masses during inflation.
We note that under these conditions, parametric resonance will not
localize the fields we are interested in close to the origin.
The case of the
$n_{\rm min}=6$ flat directions we have considered is the most
conservative, in the sense that the regions in the parameter space
which yield large scale charge and color breaking minima can be made
acceptable by lower reheat temperatures than in any of the other
$n_{\rm min}>6$ flat directions involving squarks.  The $n_{\rm min}=10$
flat directions require much higher reheating temperatures.
In general, however, different flat directions are not necessarily
mutually flat, and a large VEV in an $n_{\rm min}=6$ direction may
preclude large-scale minima in other directions.  Conservatism then
requires that we restrict our conclusions to the $n_{\rm min}=6$ case,
although there is no reason that inflation should populate
$n_{\rm min}=6$ directions in preference to directions with larger VEVs.

\vskip 0.2in
\vbox{
\noindent{ {\bf Acknowledgments} } \\
\noindent  L.R.  would like to thank D.H. Lyth for helpful discussions.
A.S. would like to thank Dan Boyanovsky,  Dan Cormier,  Hector de Vega, Prem
Sivaramakrishnan
and Rich Holman for stimulating discussions on related topics.
This work was supported in part by DOE grant DE--FG02--94ER--40823
and NSF grant PHY--91--16964.}

\newpage
\vskip 2in
\noindent{\bf{Figure Captions}}

\vskip.3truein

\begin{itemize}
 \item[]
\begin{enumerate}
\item[]
\begin{enumerate}

\item[Fig.~1)]Contours of constant $\Tc$, in units of $10^{7} \gev$,
  as a function of the gluino mass $M_3$ and the average squark mass
  $\mave=\sqrt{(\mpar^2-m_{\tilde e_R}^2)/3}$, for the $n_{\rm min}=6$
  flat direction $\tilde u_R^{\ss r} = \tilde c_R^{\ss g} = \tilde
  s_R^{\ss b} = \tilde e_R \equiv v(Q)$.  Above the top solid curve,
  the zero-temperature potential (and hence the full potential
  $V(\p,Q)+V_T$) has no second minimum.

\item[Fig.~2)]Contours of constant $\Tc$, in units of $10^{9} \gev$,
  as a function of the gluino mass $M_3$ and the average squark mass
  $\mave=\sqrt{\mpar^2/3}$, for the $n_{\rm min}=10$ flat direction
  $\tilde u_R^{\ss r} = \tilde s_R^{\ss g} = \tilde b_R^{\ss b} \equiv
  v(Q)$ .  Above the top solid curve, the zero-temperature potential
  (and hence the full potential $V(\p,Q)+V_T$) has no second minimum.

\end{enumerate}
\end{enumerate}
\end{itemize}
\newpage

\begin{figure}[htb]
\hspace{+1truecm}
\epsfysize=8.0truein
%\epsfbox{forssfig.ps}
\epsfbox{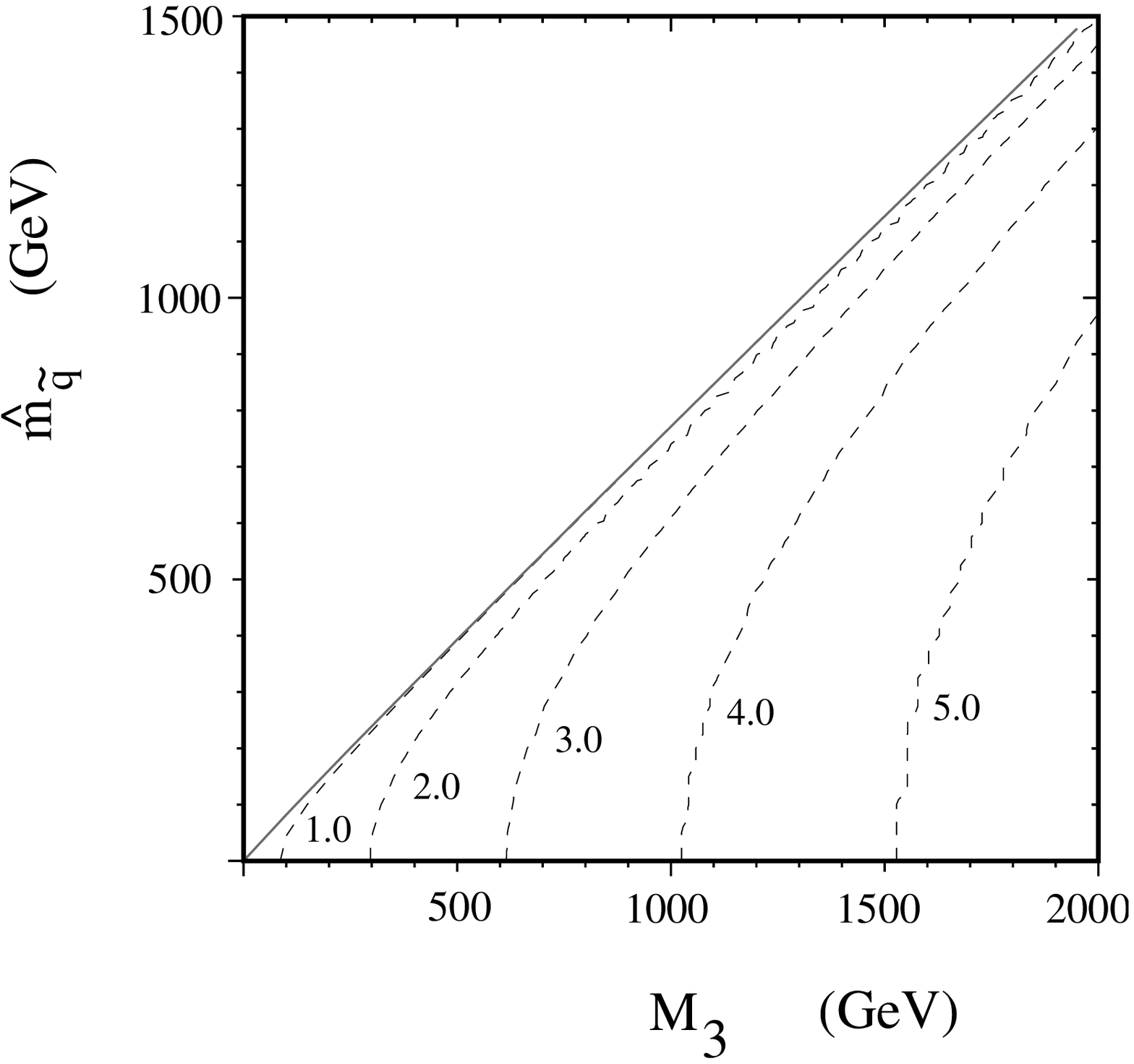}
\vspace{-0.8cm}
\begin{center} \LARGE{Figure 1} \end{center}
\end{figure}

\begin{figure}[htb]
\hspace{+1truecm}
\epsfysize=8.0truein
\epsfbox{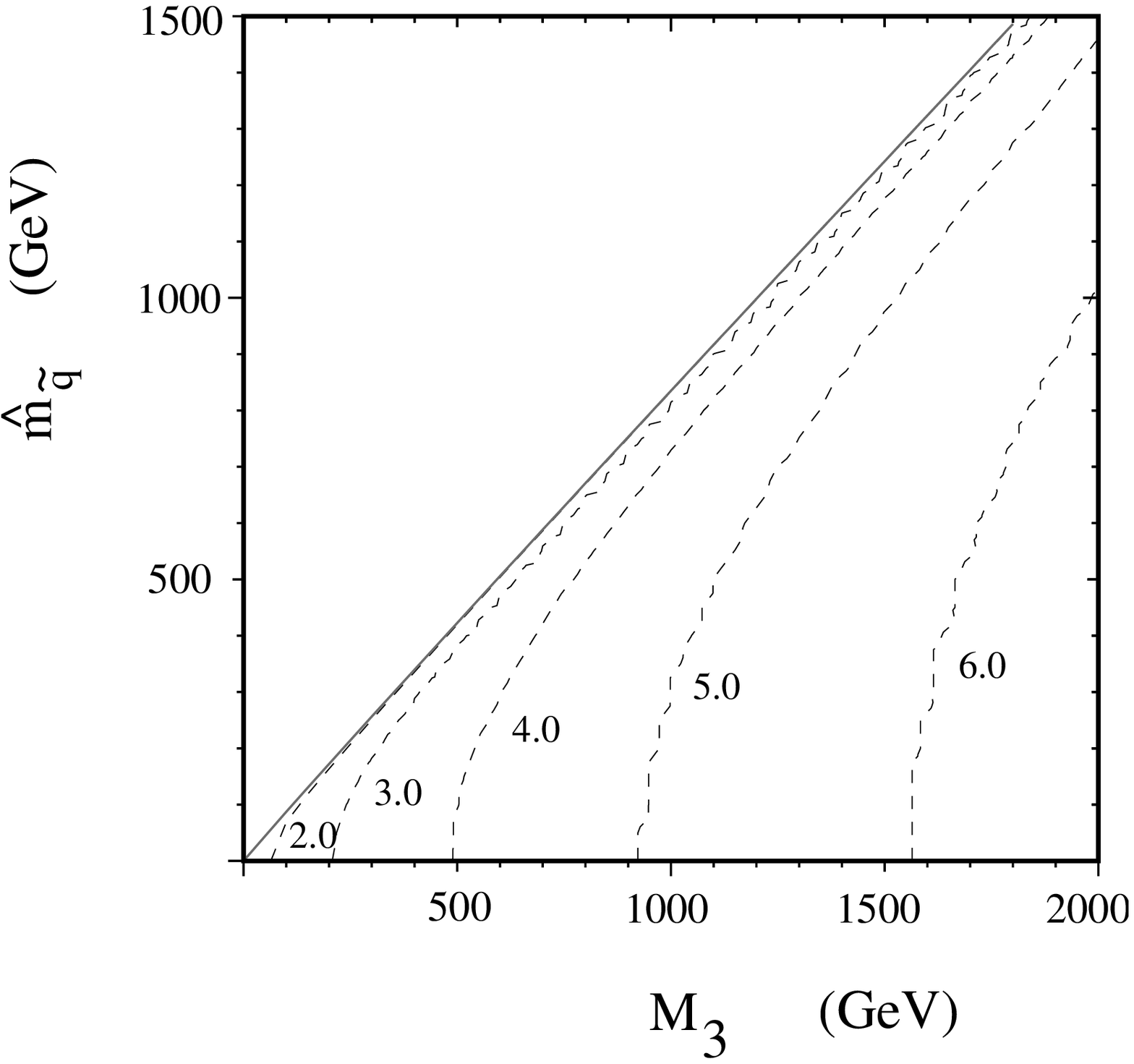}
\vspace{-0.8cm}
\begin{center} \LARGE{Figure 2} \end{center}
\end{figure}


\begin{thebibliography}{99}

\bibitem{us} T. Falk, K.A. Olive, L. Roszkowski and M. Srednicki,
Phys. Lett. {\bf B367} (1996) 183.

\bibitem{rr} A. Riotto and E. Roulet, Phys. Lett. {\bf B377} (1996) 60.

\bibitem{kls} A. Kusenko, P. Langacker, and G. Segre,
Phys. Rev. {\bf D54} (1996) 5824.

\bibitem{inf} K.A. Olive, {\it Phys. Rep.} {\bf 190} (1990) 181;
A.D. Linde, {\it Particle  Physics And Inflationary Cosmology}
(Harwood, 1990).

\bibitem{lin2}A.D. Linde, Phys. Lett. {\bf 116B} (1982) 335;
A.A. Starobinsky, Phys. Lett. {\bf 117B} (1982) 175;
A. Vilenkin, Nucl. Phys. {\bf B226} (1983) 527;
K. Enqvist, K.W. Ng, and K.A. Olive, Nucl. Phys. {\bf B303} (1988) 713.

\bibitem{cdo}B. Campbell, S. Davidson, and K.A. Olive,
Nucl.Phys. {\bf B399} (1993) 111.
\bibitem{eeno}J. Ellis, K. Enqvist, D.V. Nanopoulos, and K.A. Olive,
Phys. Lett. {\bf B191}(1987) 343.

\bibitem{ad}I. Affleck and M. Dine, Nucl. Phys. {\bf B249} (1985) 361.

\bibitem{lin3}A.D. Linde, {\it Phys. Lett.} {\bf B160} (1985) 243.

\bibitem{drt} M.Dine, L. Randall, and S. Thomas, Phys. Rev. Lett. {\bf
    75} (1995) 398; Nucl. Phys. {\bf B458} (1996) 291.
\bibitem{all}
M. Dine, W. Fischler and D. Nemeschansky, Phys. Lett. {\bf 136B} (1984) 169;
G. D. Coughlan, R. Holman, P. Ramond and G. G. Ross,
Phys. Lett. {\bf 140B} (1984) 44;
E. J. Copeland, A. R. Liddle, D. H. Lyth, E. D. Stewart and D. Wands,
Phys. Rev. {\bf D49} (1994) 6410.

\bibitem{strum} A. Strumia, hep-ph/9604417 (1996).

\bibitem{sugr} E. Cremmer, B. Julia, J. Scherk, S. Ferrara, L. Girardello and
P. Van Nieuwenhuizen, Phys. Lett. {\bf 79B} (1978) 231 and
Nucl. Phys. {\bf B147} (1979) 105;
E. Cremmer, S. Ferrara, L. Girardello and A. Van Proeyen,
Phys. Lett. {\bf 116B} (1982) 231 and
Nucl. Phys. {\bf B212} (1983) 413;
R.  Arnowitt, A.H. Chamseddine and P. Nath,
Phys. Rev. Lett. {\bf 49} (1982) 970
and {\bf 50} (1983) 232 and Phys. Lett. {\bf 121B} (1983) 33;
J. Bagger and E.  Witten, Phys. Lett. {\bf 115B} (1982) 202
and {\bf 118B} (1982) 103;
J. Bagger, Nucl. Phys. {\bf B211} (1983) 302.

\bibitem{gmo} M.-K. Gaillard, H. Murayama, and K. Olive, Phys. Lett.
{\bf B355} (1995) 71.

\bibitem{BG} P.~Binetruy and M.K.~Gaillard, 
Phys. Lett. {\bf B195} (1987) 382.

\bibitem{ns} E. Cremmer, S. Ferrara, C. Kounnas and D.V. Nanopoulos,
Phys. Lett.  {\bf 133B} (1983) 61.

\bibitem{nsguts} J. Ellis, A.B. Lahanas, D.V. Nanopoulos and K.  Tamvakis,
Phys. Lett. {\bf 134B} (1984) 429; J. Ellis, C. Kounnas and 
D.V. Nanopoulos, Nucl. Phys. {\bf B241} (1984) 406;
J. Ellis, C. Kounnas and 
D.V. Nanopoulos, Nucl. Phys. {\bf B247} (1984) 373;
for a review see: A.B. Lahanas and D.V. Nanopoulos,
Phys. Rep. {\bf 145} (1987) 1.

\bibitem{linde1} L. A. Kofman, A. Linde, and A. Starobinksy, 
Phys. Rev. Lett. {\bf 73} (1994) 3195.

\bibitem{linde2} L. A. Kofman, A. D. Linde  and A. A. Starobinsky,
Phys. Rev. Lett. {\bf 76} (1996) 1011 (1996);
I. Tkachev, Phys. Lett. {\bf B376} (1996) 35.

\bibitem{partlist} Y.~Shtanov, J.~Traschen, and R.~Brandenberger,
Phys. Rev. D {\bf 51} (1995) 5438; D. Kaiser, Phys. Rev. {\bf 53}
(1995) 1776; astro-ph/9608025; M. Yoshimura, Prog. Theor. Phys. {\bf
94}, 873 (1995); hep-th/9506176; H. Fujisaki, K. Kumekawa, M.
Yamaguchi, M. Yoshimura, Phys. Rev. D {\bf 53}, 6805 (1996); S.
Kasuya and M. Kawasaki, hep-ph/9603317; D. T. Son, hep-ph/9604340;
E.W. Kolb, A.D. Linde and A. Riotto, hep-ph/9606260; S. Khlebnikov
and I. Tkachev, Phys. Rev. Lett. {\bf 7}, 219 (1996);
S. Khlebnikov and I. Tkachev, hep-ph/9608458;  
S. Khlebnikov and I. Tkachev, hep-ph/9610477.

\bibitem{dissip} D. Boyanovsky, H. J. de Vega, R. Holman, D.-S. Lee and
A. Singh, Phys. Rev. {\bf D51}, 4419 (1995);
D. Boyanovsky, M.D'Attanasio, H. J. de Vega, R. Holman, D.-S. Lee and A. Singh,
hep-ph/9505220 (1995);
D. Boyanovsky, H. J. de Vega, R. Holman  and J. F. J. Salgado,
%Analytic and Numerical Study of Preheating Dynamics,
hep-ph/9608205;
L. Kofman, A. Linde and A. Starobinsky,
%Comments on Analytic and Numerical Study of Preheating Dynamics,
hep-ph/9608341;
D. Boyanovsky, D. Cormier, H. J. de Vega, R. Holman, A. Singh,
and M. Srednicki, hep-ph/9609527 (1996);
F. Cooper, S. Habib, Y. Kluger, E. Mottola, hep-ph/9610345.

\bibitem{prestable} G. W. Anderson, A. D. Linde and A. Riotto, 
Phys. Rev. Lett. {\bf 77} (1996) 3716;
A. Riotto, E. Roulet, I. Vilja,
%Preheating and vacuum metastability in supersymmetry,
hep-ph/9607403.

\bibitem{gher} T. Gherghetta, C. Kolda and S.P. Martin,
Nucl.Phys. {\bf B468} (1996) 37.

\bibitem{dj} L. Dolan and R. Jackiw, Phys. Rev. {\bf D9} (1974) 3320.

\bibitem{linde}A.D. Linde, Phys. Lett. {\bf B70} (1977) 306;
I. Affleck, Phys. Rev. Lett. {\bf 46} (1981) 338.

\bibitem{gw}A.H. Guth and E. Weinberg, Phys. Rev. {\bf D23} (1981)
  876.

\bibitem{grns} J. Ellis, C. Kounnas and D.V. Nanopoulos, Nucl. Phys. {\bf B247}
(1983) 373;     J. Ellis, C. Kounnas and D.V. Nanopoulos, Phys. Lett. {\bf
143B}
(1984) 410;     J. Ellis, K. Enqvist and D.V. Nanopoulos, Phys. Lett. {\bf
147B}
(1984) 99.

\bibitem{gr} H. Pagels and J. Primack,
Phys. Rev. Lett. {\bf 48} (1982) 223;
M.Y. Khlopov and A.D. Linde,
Phys. Lett. {\bf B138} (1984) 265;
D.V. Nanopoulos, K.A. Olive and
M. Srednicki, Phys. Lett. {\bf 127B} (1983) 30;
J. Ellis, J. Hagelin, D.V. Nanopoulos,
K.A. Olive and M. Srednicki,
Nucl. Phys. {\bf B238} (1984) 453;
J. Ellis, J.E. Kim and D.V. Nanopoulos,
Phys. Lett. {\bf B145} (1984) 181;
J. Ellis, D.V. Nanopoulos and S. Sarkar,
Nucl. Phys. {\bf B259} (1985) 175;
R. Juszkiewicz, J. Silk and A. Stebbins,
Phys. Lett. {\bf 158B} (1985) 463;
D. Lindley, Phys. Lett. {\bf B171} (1986) 235;
M. Kawasaki and K. Sato, Phys.
Lett. {\bf B189} (1987) 23; M. Kawasaki and T. Moroi,
Prog.Theor.Phys. {\bf 93} (1995) 879.

\bibitem{ac} R. Allahverdi and B.A. Campbell,
%Cosmological Reheating and Self-Interacting Final State Bosons
hep-ph/9606463
\end{thebibliography}
\end{document}